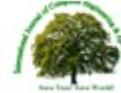

# DETERMINING THE POSSIBILITIES AND CERTAINTIES IN NETWORK PARTICIPATION FOR MANETS

**Anoop J. Sahoo[1], Md. Amir Khusru Akhtar[2]**

[1]*System Engineer, Infosys Limited, Chennaii, India*

[2] *Department of Computer Science and Engineering, Cambridge Institute of Technology, Ranchi, India*

## ABSTRACT:

A mobile ad hoc network (MANET) is a self-organized cooperative network that works without any permanent infrastructure. This infrastructure less design makes it complex compared to other wireless networks. Lot of attacks and misbehavior obstruct the growth and implementation. The majority of attacks and misbehavior can be handled by existing protocols. But, these protocols reduce the total strength of nodes in a network, because they isolate nodes from network participation having lesser reputation value.

To cope with this problem we have presented the Possibility and Certainty (P&C) model. This model uses reputation value to determine the possibilities and certainties in network participation. The proposed model classifies nodes into three classes such as 'certain' or HIGH grade, 'possible' or MED grade and 'not possible' or LOW grade. Choosing HIGH grade nodes in network activities improves the Packet Delivery Ratio (PDR) which enhances the throughput of the MANET. On the other hand when node strength is poor, we choose MED grade nodes for network activities. Thus, the proposed model allows communication in the worst scenario with the possibility of success. It protects a network from misbehavior by isolating LOW grade nodes from routing paths.

## [I] INTRODUCTION

Ad hoc is a Latin word which means 'for this task'. It is basically a solution designed for a specific task. A mobile ad hoc network (MANET) is an infrastructure-less, self-organized cooperative network in which a group of wireless devices (nodes) cooperate each other for its network operations. A wireless node may be a Personal Computer (desktops/laptops) with wireless LAN cards, Personal Digital Assistants (PDA), Palmtop or any other wireless or mobile devices. The mobile ad hoc network has many applications such as in military war zones, disaster relief operations, mine site operations and other suitable domains where infrastructures is not available, impractical, or expensive.

A MANET is a network of cooperation but when nodes misbehave all cooperation agreement fails. Nodes misbehave due to various reasons classified into honest and malicious causes [1]. In honest causes nodes want to save its valuable resources such as battery power and bandwidth. As we know energy is a scarce resource of

**(Anoop J. Sahoo and Md. Amir Khusru Akhtar)**

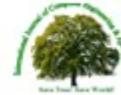

MANET and by dropping packets of other nodes wants to prolong its life. On the other hand bandwidth is another limited resource for a MANET, thus nodes drop packets of others to save its bandwidth. In malicious misbehavior nodes deploy wormhole and blackhole attacks and drop packets of others. In spite of that we have another reason for the packet drop or non cooperation such as network congestion, jamming, burst channel errors due to interference, fading, etc.

MANETs are most vulnerable to attack and misbehavior which obstruct the growth and implementation [2]. The majority of attacks and misbehavior can be handled by existing protocols. But, these existing protocols reduce the total strength of nodes in a network, because they isolate nodes from network participation having lesser reputation value. To enforce cooperation and to handle poor strength of nodes the proposed work presents the Possibility and Certainty (P&C) model which provide the extent of misbehavior that a network can allow on the basis of node strength means the number of nodes participating in the network. The proposed model uses reputation value to determine the possibilities and certainties in network participation. In order to classify nodes a network is analysed on the basis of point-valued and interval-valued methods. We have classified nodes into several classes such as 'certain' or HIGH grade, 'possible' or MED grade and 'not possible' or LOW grade. If a network having good strength of nodes, then Choosing HIGH grade nodes in network activities improves the Packet Delivery Ratio (PDR). On the other hand when node strength is poor, then the only option is to choose possible class in network activities. The proposed classification allows communication in the worst scenario with the possibility of success. Hence, the proposed P&C model enhances the throughput of the MANET as well as minimizes misbehavior by isolating LOW grade nodes from routing paths.

The rest of this paper is organized as follows: Section II presents the related work and assumption. Section III presents the Possibility and Certainty (P&C) model. Section IV discusses the experiments and results. Finally, Section V concludes the paper.

## [II] RELATED WORK AND ASSUMPTION

**(Anoop J. Sahoo and Md. Amir Khusru Akhtar)**

## 2.1. Related Work

An ad hoc network utilizes multi-hop radio relaying and operates without the support of any permanent infrastructure. A lot of work has been proposed in the literature, but they have serious limitations in terms of routing overhead and attacks. Secure routing protocols [3-8] are capable enough to handle modification of routing data, but non cooperation or misbehavior is still a challenge. Existing solutions prevent a MANET from attacks and misbehavior at some extent. But, these solutions reduce the node strength due to the stricter punishment strategy which degrades the performance of the MANET.

A lot of solutions have been proposed to prevent a network from misbehavior such as Watchdog and Pathrater [9], CONFIDANT [10-11], CORE [12] and others [13-19]. These solutions prevent a network from misbehavior and isolate misbehaved nodes from the routing paths, but these models reduce node strength due to its stricter punishment policy. Thus, these models degrade the network performance and causes network failure.

Mitrokotsa and Dimitrakakis [20] have proposed a classification algorithm for the intrusion detection in MANET. The proposed method is an innovative approach but not validated with real world data. Hernandez–Orallo et al. [21-22] have proposed the detection of selfish nodes in MANET but these algorithms consumes the valuable resources and degrade the performance. Li et al. [23] have proposed a secure routing protocol with node selfishness resistance in MANETs, but this protocol also consumes the valuable resources and degrades the performance.

## 2.2. Assumption

In this work we have taken one or more expert nodes which are responsible for storage and classification of reputation values. An expert node represents an intelligent node of the ad hoc network which has the high computation capability and capable enough to process and maintain the history of the transaction in the network [24-25]. It manages the trust and reputation values and classifies nodes of a network. The obtained classification list can be used by existing solutions [9-13] for its routing activities. The existing solutions degrade the performance of the network due to its stricter punishment strategy, but the proposed P&C model allow communication in the worst scenario

**(Anoop J. Sahoo and Md. Amir Khusru Akhtar)**

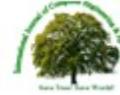

with the possibility of success. A MANET is shown in figure 1 in which Laptop node can be used as an expert node because it has the high computation capability and more battery life.

**Fig: 1. A MANET with expert node**

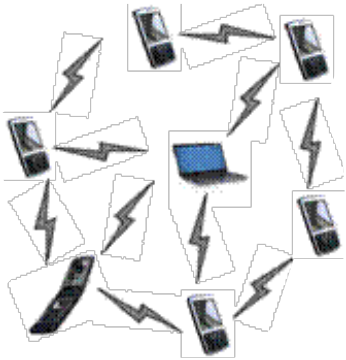

## [III] POSSIBILITY AND CERTAINTY (P&C) MODEL

This section presents the Possibility and Certainty (P&C) model. The proposed model uses reputation value to determine the possibilities and certainties in network participation and classify nodes in three classes such as 'certain' class or HIGH grade nodes, 'possible' class or MED grade nodes and 'not possible' class or LOW grade nodes. Here, Certain class means highly cooperative class, possible class means may be cooperative class and not possible class means non-cooperative nodes or misbehaving nodes. The solutions which is certain is necessarily possible because certainty implies possibility, but the converse is not always true because if something is possible it is not sure to be certain [26]. Expert nodes use reputation values of all nodes to classify and determine the possibilities and certainties in network participation. Here, reputation values are obtained from the literature [13]. In this proposed model a network is analysed and classified using point-valued and interval-valued methods [26]. Table 1 shows the grade, class and usage of the classification.

| Grade | Class | Usage |
|-------|-------|-------|
| HIGH | certain | If network activities are performed through these nodes then the communication is certain. |
| MED | possible | If network activities are performed through these nodes then the communication is possible. |
| LOW | not possible | If network activities are performed through these nodes then the communication is not possible. |

**Table: 1.** Grade, Class and Usage of classification

### 3.1. Point-valued method

In point-valued method reputation value is compared between the defined query range to know the network participation. Let us consider the defined query range is Q (x,y) and we denote the reputation values R of the $i^{th}$ node as $R_i$. The query range is defined on the basis of network scenario means how much cooperation is desired

**(Anoop J. Sahoo and Md. Amir Khusru Akhtar)**

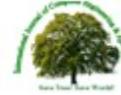

for the network in the current situation. In this work we have taken the query range on the basis of number of nodes participating in a network, because node strength is a major requirement for a cooperative network. Here, we have taken reputation values between 0 to 100 as defined in [13]. The minimum and maximum reputation values are denoted by MINR and MAXR respectively. The network participation of nodes in point-valued method is classified using the given function.

```
#define MINR 0
#define MAXR 100
void point_valued_method (R_i, x,y)
{
if ((R_i >= x && R_i >= y && R_i <=MAXR))
        Grade ("HIGH");
    else if (R_i >= x)
        Grade ("MED");
    else if (R_i > MINR && R_i <x)
        Grade ("LOW");
    else
        Grade ("Reputaion value error");
}
```

**3.2. Interval-valued or set-valued method**

Now consider a scenario in which nodes reputation values are not known precisely, but rather each node reputation value defined as an interval. These reputation values are obtained in different time units of a simulation experiment and maintained by the expert nodes. In this case reputation values are set-valued quantities. Let us consider the defined query range is Q (x,y) and we denote the reputation values R of the $i^{th}$ node as $R_i(p,q)$. The network participation of nodes in interval-valued method is classified using the given function.

```
#define MINR 0
#define MAXR 100
void interval_valued_method (p,q, x,y)
{
    if ((p>=x && p<= y) &&
      (q>=x && q <= MAXR))
            Grade ("HIGH");
    else if ((p>=x && p<= y) ||
          (q>=x && q <= Y))
                Grade ("MED");
    else if (p < MINR || q < MINR)
                Grade ("LOW");
    else
        Grade ("Reputaion value error");
```

**(Anoop J. Sahoo and Md. Amir Khusru Akhtar)**

}

### 3.3. Application of the proposed P&C model

The proposed classification is suitable when a network is measured on the basis of its size, the number of nodes participating in the network. Table 2 shows the application of the proposed model. It shows that if the size of the network is measured in terms of number of nodes, then choosing HIGH grade nodes or certain class improves the Packet Delivery Ratio (PDR) as well as increases the throughput of the MANET. On the other hand when the network having a limited number of nodes, then the only option is to choose MED grade nodes or possible class. Thus, the proposed classification allows communication in the worst scenario with the possibility of success. But, LOW grade nodes should be isolated from network activities because, they are malicious or non-cooperative.

| Network size | Grade | Class | Application |
|---|---|---|---|
| Large | HIGH | certain | always allowed in network activities |
| Small | MED | possible | allowed in network activities when node strength is limited |
| Large or Small | LOW | not possible | isolated from network activities |

**Table: 2.** Application of P&C model

## [IV] EXPERIMENTS AND RESULTS

In order to obtain the results several experiments are performed on the basis of the point-valued and Interval-valued methods. In this work reputation values are taken in the range of 0 to 100 as defined in [13]. We have classified the network into HIGH, MED and LOW grade nodes using point-valued and Interval valued method.

Let us consider a network of of nine nodes in which reputation values are taken in terms of point-valued and interval-valued quantities. Figure 2 shows proposed network in which nodes are at a distance of one hop from expert node and arranged in a grid pattern, where arrows indicate bidirectional flow.

**Fig: 2. Proposed network**

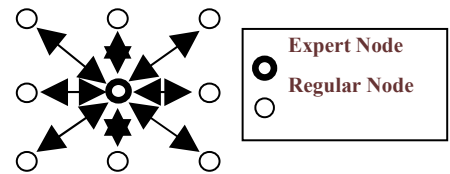

The classification using the proposed P&C model is as follows:

### 4.1. Classification using Point-valued method

Let us consider the defined query range is Q (50, 70). On the basis of the network

**(Anoop J. Sahoo and Md. Amir Khusru Akhtar)**

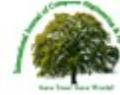

information provided in the Table 3 nodes can be classified into three classes.

| Node_ID | Reputation_Value |
|---------|------------------|
| 1 | 46 |
| 2 | 56 |
| 3 | 90 |
| 4 | 78 |
| 5 | 33 |
| 6 | 24 |
| 7 | 56 |
| 8 | 78 |

**Table: 3.** Point-valued quantities

The classification on the basis of the Point-valued method is shown in Table 4.

| Node_ID | Class |
|---------|-------|
| 1 | Not possible |
| 2 | possible |
| 3 | certain |
| 4 | certain |
| 5 | Not possible |
| 6 | Not possible |
| 7 | possible |
| 8 | certain |

**Table: 4.** Classification using Point-valued method

On the basis of Table 4 we can classify nodes into three classes. The 'certain' class having nodes {3, 4, 8}, in 'possible' class we have {2, 7} and nodes {1, 5, 6} are in 'not possible' class. Choosing certain class in network activities improves the Packet Delivery Ratio (PDR) as well as increases the throughput of the MANET. On the other hand we have limited number of nodes in the network then we choose 'possible' class for network activities. But, not possible class should be isolated from network activities because, they are malicious or non-cooperative.

### 4.2. Classification using Interval-valued method

Now consider a scenario in which nodes reputation values are not known precisely, but rather each node reputation value defined as an interval. These reputation values are obtained in five time units of a simulation experiment. In this case reputation values are set-valued quantities, as given in Table 5.

| Node_ID | Reputation_Value |
|---------|------------------|
| 1 | 46-60 |
| 2 | 56-70 |
| 3 | 90-95 |
| 4 | 78-90 |
| 5 | 33-45 |
| 6 | 24-50 |
| 7 | 56-60 |
| 8 | 78-80 |

**Table: 5.** Interval-valued quantities

On the basis of the set-valued quantities we can classify nodes using the interval-valued method, because the reputation values of the nodes are expressed in terms of ranges as well as the solution space are also expressed in terms of ranges.

Let us consider the defined query range is Q (50, 70). The classification on the basis of Interval-valued method is shown in Table 6.

**(Anoop J. Sahoo and Md. Amir Khusru Akhtar)**

| Node_ID | Reputation_Value |
|---------|------------------|
| 1 | possible |
| 2 | certain |
| 3 | certain |
| 4 | certain |
| 5 | not possible |
| 6 | possible |
| 7 | certain |
| 8 | certain |

**Table: 6.** Classification using Interval-valued method

From Table 6 it is clear that {2, 3, 4, 7, 8} are in 'certain' class, {1, 6} are in 'possible' class, {5} is in 'not possible' class. Choosing 'certain' class in network activities improves the Packet Delivery Ratio (PDR) as well as increases the throughput of the MANET. On the other hand when the network having a limited number of nodes then we can choose 'possible' class for network activities. But, 'not possible' class should be isolated from network activities because node 5 is malicious or non-cooperative.

## [V] CONCLUSION

A mobile ad hoc network is a cooperative network in which nodes have dual tasks of forwarding and routing thus, network participation of all nodes is highly desirable. The proposed model classify nodes in certain, possible and not possible classes on the basis of reputation values. The proposed classification is suitable when a network is measured on the basis of its size means number of nodes participating in the network. Choosing HIGH grade nodes or 'certain' class in network activities improves the Packet Delivery Ratio (PDR). On the other hand when a network having limited number of nodes, then the proposed work chooses MED grade nodes or 'possible' class in network activities. Thus, allow communication in the worst scenario with the possibility of success. In this work LOW grade nodes should be isolated from network activities because, they are malicious or non-cooperative. The obtained result in section IV shows the efficiency of the model. Hence, the proposed P&C model enhances the throughput of the MANET as well as minimizes misbehavior by isolating LOW grade nodes.

**(Anoop J. Sahoo and Md. Amir Khusru Akhtar)**

**(Anoop J. Sahoo and Md. Amir Khusru Akhtar)**

**(Anoop J. Sahoo and Md. Amir Khusru Akhtar)**